\documentclass[11pt,a4paper]{amsart}
\usepackage{amssymb}
\usepackage[final]{showkeys}
\usepackage{microtype}
\usepackage{color}
\usepackage{graphicx}
\usepackage[hmargin=3cm,vmargin={3.5cm,4cm}]{geometry}
\numberwithin{equation}{section}

\allowdisplaybreaks

\begin{document}

    \title[Non-homogeneous nonlinear diffusion equations]{On the relation between non-homogeneous
    fractional Burgers equations and
    time-dependent harmonic oscillator.}

    \author{P. Artale Harris$^1$}
     \address{${}^1$ Istituto Nazionale di Geofisica e Vulcanologia, Roma.}

    \author{R. Droghei$^{2}$}
     \address{${}^2$IIS-Ceccano, Via Gaeta, 105, 03023 Ceccano FR.}
     
    \author{R. Garra$^3$}
    \address{${}^3$Dipartimento di Scienze Statistiche, ``Sapienza'' Universit\`a di Roma.}

    \author{E. Salusti$^4$}
    \address{${}^4$ CNR-ISMAR, Via Fosso del Cavaliere 100, 00133 Roma.}


    \begin{abstract}
    In this paper we discuss the relation between non-homogeneous nonlinear fractional diffusive
    equations and the Schr\"odinger equation with time-dependent
    harmonic potential. It is well known that the
    Cole-Hopf transformation allows to linearize non-homogeneous nonlinear diffusive equations (NHNDEs)
    into a Schr\"odinger-type equation with time-dependent
    potential. We first discuss the utility of the results about
    time-dependent harmonic oscillator to build explicit solutions
    of such non-homogeneous nonlinear partial differential equations. In
    particular, we recall that starting from a trial polynomial
    solution of the NHNDE, it is possible to construct other
    solutions by using linear invariants of the Schr\"odinger equation with time-dependent
    potential. Finally we apply these results to find explicit
    solutions to a novel non-homogeneous fractional Burgers-type
    equation.
    \end{abstract}

\maketitle
\section{Introduction}
One of the most known example of linearizable nonlinear partial
differential equations is the classical Burgers equation, firstly
introduced by J.M Burgers in 1948 in the framework of the theory of
turbulence \cite{Burgers}. In recent years several studies have been
devoted to the analysis of different generalizations of Burgers-type
equations. In particular, two kind of generalizations have been
widely studied in literature: non-homogeneous Burgers equation and
generalized Burgers equation with time-dependent viscosity and
drift. In the framework of non-homogeneous Burgers equations, Rao
and Yadav \cite{Rao} considered the following equation
\begin{equation}\label{Raos}
\frac{\partial u}{\partial t}+u\frac{\partial u}{\partial
x}-\frac{\partial^2 u}{\partial x^2}=\frac{kx}{(2\beta t+1)^2},\quad
x\in \mathbb{R}, t>0,
\end{equation}
subject to some unbounded initial conditions. The solutions of
\eqref{Raos} are expressed in terms of the self-similar solutions of
a linear partial differential equation with variable coefficients.
In this way the large time behavior of the solutions is obtained. A
similar equation was considered by Eule and Friedrich \cite{Eule}
with a more general non-homogeneous term, again linear in space.
 More
recently, exact solutions of forced Burgers equation with
time-variable coefficients have been discussed by Buyukasik and
Pashaev in \cite{tur}. In \cite{sch}, Darboux transformations have
been applied to the time-dependent non-homogeneous Burgers equation,
in relation with the Schr\"odinger equation. Finally an interesting
relation between the Burgers equation with elastic forcing and the
Ornstein-Uhlenbeck process
was pointed out in \cite{Moreau}.\\
In the framework of generalized Burgers equation with time-dependent
viscosity, Rao and Satyanarayana in \cite{Rao2} discussed the
large-time asymptotics for solutions of Burgers equation with
variable viscosity related to nonlinear acoustics. Vaganan and
Jeyalakshmi \cite{vaga} considered the following more general
Burgers equations with linear damping and variable viscosity
\begin{equation}\label{drift}
\frac{\partial u}{\partial t}+u\frac{\partial u}{\partial
x}-\frac{\Delta(t)}{2}\frac{\partial^2 u}{\partial x^2}+\alpha u=0,
\quad \alpha >0,
\end{equation}
discussing its relation with several linear parabolic equations. In
this paper the authors discuss conditions on the viscosity
time-dependence $\Delta(t)$ for which
\eqref{drift} is linearizable.\\

Here we consider the following non-homogeneous nonlinear diffusive
equation (NHNDE)
\begin{equation}
\frac{\partial u}{\partial t}+\left(\frac{\partial u}{\partial
x}\right)^2-\frac{\partial^2 u}{\partial x^2}=V(x,t). \label{nhdes}
\end{equation}
This NHNDE is widely used in different fields of applied sciences.
For example similar nonlinear diffusive equations arise in the
analysis of the propagation of nonlinear waves in porous media (see
e.g. \cite{pippo1} and references therein). It is easy to show that
such equation is related to the non-homogeneous Burgers equation.
Indeed it suffices to differentiate both sides in \eqref{nhdes} in
order to obtain
\begin{equation}
\frac{\partial}{\partial t}\frac{\partial u}{\partial
x}+2\frac{\partial u}{\partial x}\frac{\partial^2 u}{\partial
x^2}-\frac{\partial^3 u}{\partial x^3}=\frac{\partial
V(x,t)}{\partial x},
\end{equation}
that coincides with the non-homogeneous Burgers equation for
$f(x,t)=\frac{\partial u}{\partial x}$.\\
For a suitable non-homogeneous term $V(x,t)$ (see \eqref{pote}
behind), we discuss the relation between NHNDE and time-dependent
harmonic oscillator. In more detail we underline the utility to
apply known results about Hamiltonian invariants of the
time-dependent harmonic oscillator in order to construct exact
solutions of NHNDEs, starting from polynomial-type trial solutions.
This approach can be applied to handle more general NHNDE with
variable viscosity of the form
\begin{equation}
\frac{\partial u}{\partial t}+\left(\frac{\partial u}{\partial
x}\right)^2-\frac{1}{2m(t)}\frac{\partial^2 u}{\partial x^2}=V(x,t),
\end{equation}
under suitable conditions on the given functions $V(x,t)$ and
$m(t)$.

The aim of this paper is to apply these results to solve a novel
class of non-homogeneous nonlinear nonlocal diffusive equations,
that is
\begin{eqnarray}\label{misk}
\qquad\frac{\partial u}{\partial t}+\frac{\partial^\alpha}{\partial
x^\alpha}\left[\left( \frac{\partial^{1-\alpha }u}{\partial
x^{1-\alpha}}\right)^2 - \frac{\partial^2}{\partial
x^2}\int_0^x\frac{(x-x')^{\alpha-1}}{\Gamma(\alpha)}u(x')dx'\right]
=-\frac{k(t)x^{2-\alpha}}{\Gamma(3-\alpha)},
\end{eqnarray}
where
\begin{equation}
\int_0^x\frac{(x-x')^{\alpha-1}}{\Gamma(\alpha)}u(x',t)dx'=
J_x^{\alpha}u(x,t),
\end{equation}
is the Riemann-Liouville fractional integral of order $\alpha \in
(0,1]$ and
\begin{equation}
\frac{\partial^\alpha}{\partial x^\alpha} u(x,t)=
\frac{1}{\Gamma(1-\alpha)}\int_0^{x}(x-x')^{-\alpha}\frac{\partial}{\partial
x'}u(x',t) dx',
\end{equation}
is the Caputo fractional derivative with respect to $x$-variable
(see e.g. \cite{pod, kil} and Appendix A for more details). This is
a slight modification of the Burgers equation considered by Miskinis
in \cite{mi1, mi2}. As far as we know, this is the unique case
discussed in literature, where a nonlinear nonlocal differential
equation can be reduced to the linear heat equation by means of a
generalized Cole-Hopf transform. We remark that equation
\eqref{misk} for $\alpha = 1$ is the Burgers equation with elastic
forcing. Even if this formulation of fractional Burgers equation
seems to be in some way artificial, it has the great advantage and
interest to couple nonlinear and nonlocal effects in the dynamics.

The paper is organized as follows. In sections 2 and 3 we discuss
the relations between non-homogeneous nonlinear diffusion equations
and time-dependent harmonic oscillator, showing the utility of
invariant analysis to find explicit solutions. In section 4 we apply
the discussed results to solve a non-homogeneous fractional
Burgers-type equation.

\section{Relations between non-homogeneous
    nonlinear diffusion equations and
    time-dependent harmonic oscillator}

Let us consider the NHNDE
\begin{equation}
\frac{\partial u}{\partial t}+\left(\frac{\partial u}{\partial
x}\right)^2-\frac{\partial^2 u}{\partial x^2}=V(x,t). \label{nhde}
\end{equation}
 In order to linearize the
previous equation we use the Cole-Hopf transformation
\begin{equation}
u(x,t)=-\log\psi(x,t). \label{cole}
\end{equation}
By substituting \eqref{cole} in \eqref{nhde}, we obtain
\begin{equation}
\frac{\partial \psi}{\partial t}-\frac{\partial^2 \psi}{\partial
x^2}=-V(x,t)\psi. \label{linear}
\end{equation}
From now on, we assume that
\begin{equation}\label{pote}
V(x,t)=\frac{1}{4}\omega^2k(t)\hbar^2 x^2,
\end{equation}
and with the transformation
\begin{eqnarray}
\nonumber& \tau=i\hbar t,\\
& \xi=\frac{i\hbar}{\sqrt{2m}}x,
\end{eqnarray}
the Schr\"odinger equation with time-dependent harmonic potential is
obtained
\begin{equation}\label{sc}
i\hbar\frac{\partial\psi}{\partial
\tau}=-\frac{\hbar^2}{2m}\frac{\partial^2\psi}{\partial
\xi^2}+\frac{1}{2}m\omega^2\xi^2k(\tau)\psi.
\end{equation}
We will now follow the argument of  Salusti and Zirilli in
\cite{sal}, where in particular the authors have proven that a trial
solution for \eqref{sc} of the form
\begin{equation}
\psi(\xi,\tau)=\exp\left[\frac{i}{2\hbar}(a(\tau)\xi^2+2b(\tau)\xi+
c(\tau))\right]
\end{equation}
can be obtained. This leads to the following system
\begin{eqnarray}
\begin{cases}
\dot{a}=-\frac{1}{m}a^2-m\omega^2k\\
\dot{b}=-\frac{1}{m}ab\\
\dot{c}=\frac{i\hbar}{m}a-\frac{1}{m}b^2.
\end{cases}
\end{eqnarray}
Going back to the original problem, i.e. by using the transformation
\eqref{cole}, we obtain a solution of the non-homogeneous nonlinear
diffusion equation \eqref{nhde} in the form
\begin{equation}
u(\xi, \tau)=\frac{i}{2\hbar}(a(\tau)\xi^2+2b(\tau)\xi+ c(\tau)),
\end{equation}
and finally, going back to the original variables $(x,t)$, we have a
polynomial solution of equation \eqref{nhde}. The existence of this
kind of solutions with separate variables is not surprising. Indeed
the same relation can be obtained from the Invariant Subspace
Method, introduced by Galaktionov \cite{Gala},
which allows to solve exactly nonlinear equations by separating variables.\\
We recall the main idea of this method: consider a scalar evolution
equation
\begin{equation}\label{pro}
\frac{\partial u}{\partial t}= F\left[u, \frac{\partial u}{\partial
x}, \dots\right],
\end{equation}
where $u=u(x,t)$ and $F[\cdot]$ is a nonlinear differential
operator. Given $n$ linearly independent functions
$$f_1(x), f_2(x),....,f_n(x),$$
we call $W_n$ the $n$-dimensional linear space
$$W_n=\langle f_1(x), ...., f_n(x)\rangle.$$
This space is called invariant under the given operator $F[u]$, if
$F[y]\in W_n$ for any $y\in W_n$. This means that there exist $n$
functions $\Phi_1, \Phi_2,..., \Phi_n$ such that
$$F[C_1f_1(x)+......C_n f_n(x)]= \Phi_1(C_1,....,C_n)f_1(x)+......+\Phi_n(C_1,....,C_n)f_n(x),$$
where $C_1, C_2, ....., C_n$ are arbitrary constants. \\
Once the set of functions $f_i(x)$ that form the invariant subspace
has been determined, we can search an exact solution of \eqref{pro}
in the invariant subspace in the form
\begin{equation}
u(x,t)=\sum_{i=1}^n u_i(t)f_i(x).
\end{equation}
where $f_i(x)\in W_n$. In this way, we arrive to a system of ODEs.
In many cases, this problem is simpler than the original one and
allows to find exact
solutions by just separating variables \cite{Gala}.\\
We can now show by direct calculations that \eqref{nhde} admits $W_3
=\langle 1,x,x^2\rangle$ as invariant subspace. Indeed, we have that
\begin{equation}
F\left[u, \frac{\partial u}{\partial x}, x,
t\right]=-\left(\frac{\partial u}{\partial
x}\right)^2+\frac{\partial^2 u}{\partial x^2}
-\frac{1}{4}\omega^2k(t)\hbar^2 x^2,
\end{equation}
so that
\begin{equation}
F[c_1 x^2+ c_2 x+ c_3]=-4c_1^2 x^2-c_2^2-4c_1c_2x+2c_1
-\frac{1}{4}\omega^2k(t)\hbar^2 x^2,
\end{equation}
as claimed.

\subsection{Linear Invariants}
In the algebraic approach, we want to find an operator $\hat{A}(t)$
such that, if $\psi(x,t)$ is a solution of equation \eqref{linear},
also $\hat{A}\psi(x,t)$ is a solution. Recalling that \eqref{sc}, is
related to the Hamiltonian
\begin{equation}
\hat{H}=-\frac{\hbar^2}{2m}\frac{\partial^2}{\partial
\xi^2}+\frac{1}{2}m \omega^2 k(t) \xi^2,
\end{equation}
we want to determine an operator $\hat{A}$ such that
\begin{equation}\label{inva}
i\hbar \frac{\partial}{\partial
t}\hat{A}\psi(x,t)=\hat{H}\hat{A}\psi(x,t).
\end{equation}
The definition of invariants, according to \eqref{inva}, was firstly
introduced by Lewis and Riesenfeld in \cite{Lewis} and gave rise to
a number of papers devoted to the analysis and application of the
Schr\"odinger equation with time-dependent
    harmonic potential.
This requirement leads to the following equation (see e.g.
\cite{sal})
\begin{equation}
i\hbar\dot{\hat{A}}=[\hat{H},\hat{A}]. \label{commute}
\end{equation}
Salusti and Zirilli have shown that the Ansatz
\begin{equation}
\hat{A}=\alpha(t)x+\beta(t)\frac{\partial}{\partial x}+\gamma(t)
\end{equation}
satisfies equation \eqref{commute}. This argument leads to the
system
\begin{eqnarray}
\begin{cases}
i\hbar\dot{\alpha}=-m\omega^2 k\beta\\
i\hbar\dot{\beta}=-\frac{\hbar^2}{m}\alpha\\
\dot{\gamma}=0.
\end{cases}
\label{system2}
\end{eqnarray}
We can decouple system \eqref{system2} by deriving the second
equation with respect to $t$ and, by substitution, we obtain
\begin{equation}\label{H}
\ddot{\beta}=-k(t)\omega^2\beta,
\end{equation}
that is a time-dependent harmonic oscillator. \\
We conclude that, once fixed the function $k(t)$ in \eqref{sc}, we
can solve equation \eqref{H} and thus by \eqref{system2}, we obtain
the explicit form of the operator $\hat{A}$. Finally we can find
other solutions to the non-homogeneous nonlinear \eqref{nhde}. We
remark that this procedure can be applied in an iterative way to
construct
a family of exact solutions starting from the simple polynomial one.\\
For instance by taking $k(t)=t$, we can find an explicit solution in
terms of a cumbersome combination of Airy functions, as can be seen
by using simple mathematical routines.

\section{Non-homogeneous nonlinear diffusive equation with variable coefficients}
We now consider the equation
\begin{equation}
\frac{\partial u}{\partial t}+\left(\frac{\partial u}{\partial
x}\right)^2-\frac{1}{2m(t)}\frac{\partial^2 u}{\partial x^2}=V(x,t),
\label{new}
\end{equation}
where $V(x,t)$ has the form
\begin{equation}\label{new1}
V(x,t)=\frac{1}{4}\omega^2 m(t) k(t)\hbar^2 x^2,
\end{equation}
and $m(t)$ is a time-dependent ''good'' function. This equation is
strictly related to the one studied in \cite{tur}. Indeed it can be
reduced to the Burgers equation with variable coefficients and
elastic forcing with time-dependent elasticity.\\
By using again the transformation
\begin{equation}
u(x,t)=-\log\psi(x,t),\label{new2}
\end{equation}
we get
\begin{equation}
\frac{\partial \psi}{\partial t}-\frac{1}{2m(t)} \frac{\partial^2
\psi}{\partial x^2}=-\frac{1}{4}\omega^2 m(t) k(t)\hbar^2 x^2\psi.
\label{new3}
\end{equation}
Using the transformation
\begin{eqnarray}
\nonumber& \tau=i\hbar t,\\
& \xi=i\hbar x,
\end{eqnarray}
we obtain in this case, the Schr\"odinger equation with
time-dependent harmonic potential and variable in time masses, that
is
\begin{equation}\label{new4}
i\hbar\frac{\partial\psi}{\partial
\tau}=-\frac{\hbar^2}{2m(t)}\frac{\partial^2\psi}{\partial
\xi^2}+\frac{1}{2}m(t)\omega^2\xi^2k(\tau)\psi.
\end{equation}
Also in this case, we can find an explicit solution by using the
Gaussian Ansatz
\begin{equation}
\psi(\xi,\tau)=\exp\left[\frac{i}{2\hbar}(a(\tau)\xi^2+2b(\tau)\xi+
c(\tau))\right].
\end{equation}
In this case we find by substitution, that the functions $a(\tau)$,
$b(\tau)$ and $c(\tau)$ must satisfy the following system of ODEs
\begin{eqnarray}
\begin{cases}
\dot{a}=-\frac{1}{m}a^2-m\omega^2k\\
\dot{b}=-\frac{1}{m}ab\\
\dot{c}=\frac{i\hbar}{m}a-\frac{1}{m}b^2.
\end{cases}
\end{eqnarray}
This leads to find a polynomial solution to the original problem
\eqref{new}. We now apply the result discussed by Bassetti et al.
\cite{bassetti} in order to find also in this case other solutions
starting from the basic polynomial one. In this case the linear
invariant of the form
\begin{equation}
\hat{A}=\alpha(t)x+\beta(t)\frac{\partial}{\partial x}+\gamma(t)
\end{equation}
leads to the system
\begin{eqnarray}
\begin{cases}\label{ciao}
i\hbar\dot{\alpha}=-m(t)\omega^2 k(t)\beta(t)\\
i\hbar\dot{\beta}=-\frac{\hbar^2}{m(t)}\alpha(t)\\
\dot{\gamma}=0.
\end{cases}
\end{eqnarray}
By decoupling system \eqref{ciao}, we obtain
\begin{equation}
\frac{1}{2m(t)}\frac{d}{dt}\left({\beta (t)}-\frac{1}{2
m(t)}\frac{d}{dt}{\beta (t)}\right)+\frac{\omega^2 \beta(t)
k(t)}{4m(t)}=0
\end{equation}
In this case the study is more complicated, because of the
time-dependence of mass and elasticity. However once fixed the form
of these functions, it is possible to find a general solution by
using the invariants as in the previous section.

\section{The non-homogeneous nonlinear fractional diffusive equation}
As mentioned in the introduction, we now apply the results discussed
in the previous sections in order to solve the following
non-homogeneous nonlinear fractional Burgers equation
\begin{eqnarray}
\frac{\partial u}{\partial t}+\frac{\partial^\alpha}{\partial
x^\alpha}\left[\left( \frac{\partial^{1-\alpha }u}{\partial
x^{1-\alpha}}\right)^2 - \frac{\partial^2}{\partial
x^2}J_x^{\alpha}u\right]
=-\frac{k(t)x^{2-\alpha}}{\Gamma(3-\alpha)}, \label{miskinis}
\end{eqnarray}
where the fractional derivatives are in the sense of Caputo (see Appendix A) and $\alpha \in (0,1]$.
For $\alpha = 1$ this is the Burgers equation with elastic forcing.
We study a class of explicit solutions of this equation, considering the problem in the semi-line $x\geq 0$ with an appropriate boundary condition
in $x=0$ that will be discussed in what follows.   
Note that the in the right hand side term we can write
\begin{eqnarray}
\frac{k(t)}{2}\frac{\partial^\alpha x^2}{\partial x^\alpha}=
\frac{k(t)x^{2-\alpha}}{\Gamma(3-\alpha)}
\end{eqnarray}
so that equation \eqref{miskinis} becomes
\begin{eqnarray}
\frac{\partial u}{\partial t}+\frac{\partial^\alpha}{\partial
x^\alpha}\left[\left( \frac{\partial^{1-\alpha }u}{\partial
x^{1-\alpha}}\right)^2 -\frac{\partial^\alpha}{\partial x^\alpha}
\frac{\partial^2}{\partial x^2}J^\alpha_x u+
\frac{1}{2}k(t)x^2\right]=0. \label{miskinis2}
\end{eqnarray}
Now consider the following Cole-Hopf transoformation
\begin{eqnarray}\label{copf}
u(x,t)=-\frac{\partial^\alpha}{\partial x^\alpha}\log\psi(x,t).
\end{eqnarray}
We now observe that, by definition of Caputo and Riemann-Liouville fractional derivatives, the following equality holds 
\begin{equation}
\frac{\partial^{1-\alpha }}{\partial
x^{1-\alpha}}\frac{\partial^{\alpha}u }{\partial x^{\alpha}} = 
J_x^{\alpha}\frac{\partial}{\partial x}J_x^{1-\alpha}\frac{\partial u}{\partial x}= J_x^{\alpha}\frac{{}^R \partial}{\partial x^{\alpha}}
\frac{\partial u}{\partial x}.
\end{equation}
Moreover it can be proved that
\begin{equation}
J_x^{\alpha} \frac{{}^R \partial}{\partial x^{\alpha}} u(x,t)= u(x,t)-u(0,t), \qquad \alpha \in (0,1], \: x>0
\end{equation}
Considering these preliminaries, we now assume that 
$\ln\psi(x,t)\bigg|_{x=0}= \partial_x \ln \psi(x,t)\bigg|_{x=0}=0$.
This is equivalent to take $u(0,t)=0$ as boundary condition.
Under these assumption, we have
\begin{eqnarray}
\frac{\partial^\alpha}{\partial x^\alpha}\left[-\frac{\partial_t\psi}{\psi}+\left(\frac{\partial_x\psi}{\psi}\right)^2-\left(\frac{\partial_x\psi}{\psi}\right)^2+\frac{\partial_{xx}\psi}{\psi}\right]=-\frac{1}{2}k(t)\frac{\partial^\alpha x^2}{\partial x^\alpha},
\end{eqnarray}
thus,
\begin{eqnarray}
\frac{\partial^\alpha}{\partial x^\alpha}\left[-\frac{\partial_t\psi}{\psi}+\frac{\partial_{xx}\psi}{\psi}+\frac{1}{2}k(t)x^2\right]=0.
\end{eqnarray}
This means that
\begin{eqnarray}
\frac{\partial_t\psi}{\psi}-\frac{\partial_{xx}\psi}{\psi}-\frac{1}{2}k(t)x^2=f(t).
\end{eqnarray}
Now, suppose $f(t)=0$, we arrive to the following equation
\begin{eqnarray}\label{miskinisa}
\frac{\partial_t\psi}{\psi}-\frac{\partial_{xx}\psi}{\psi}=\frac{1}{2}k(t)x^2
\label{linear2}
\end{eqnarray}
that is the analogous of equation \eqref{linear}. This means that we
can apply exactly the same reasoning discussed in the previous
sections in order to construct exact solutions for the equation
\eqref{miskinis} starting from a polynomial trial solution. Indeed,
we can start again from the Gaussian ansatz
\begin{eqnarray}
\psi(x,t)=\exp[a(t)x^2+b(t)x+c(t)],
\end{eqnarray}
as a trial solution for \eqref{miskinisa}. Then, by using
\eqref{copf} we obtain the following solution of equation
\eqref{miskinis2}
\begin{eqnarray}\label{misks}
u(x,t)&=&-\frac{\partial^\alpha}{\partial
x^\alpha}(a(t)x^2+b(t)x+c(t))\nonumber\\&=&-
\left(\frac{b(t)x^{1-\alpha}}{\Gamma(2-\alpha)}+\frac{2a(t)x^{2-\alpha}}{\Gamma(3-\alpha)}\right),
\end{eqnarray}
where $a(t)$ and $b(t)$ satsify the following system of differential equations
\begin{eqnarray}\label{sys}
&\dot{a}=4a^2+\frac{k(t)}{2},\\
\nonumber &\dot{b}=4ab,
\end{eqnarray}
whose solutions clearly depend by the 
particular choice of $k(t)$. Then, by using again the relations with
the time-dependent Schr\"odinger equation and the analysis of
invariants, it is possible to find other explicit solutions starting
from \eqref{misks}.

We finally observe that the solution \eqref{misks} corresponds to
the fact that the equation \eqref{miskinis} admits as invariant
subspace $W_2 =\langle x^{1-\alpha},x^{2-\alpha}\rangle$, for all
$\alpha \in (0,1]$.

\subsection{The case $k(t)=t$}
We solve the system of differential equations (\ref{sys}) considering the particular choice $k(t)=t$. The first equation of the system become:

\begin{equation}
\dot a=4a^2+\frac{t}{2}
%u_2=\frac{1}{2\Gamma(3-\alpha)}\frac{\phi'}{\phi}
\end{equation}
that is a Special Riccati equation with solution

\begin{equation}\label{ch}
%\dot a=4a^2+\frac{t}{2}
a(t)=\frac{1}{4}\frac{\dot \phi(t)}{\phi(t)}
\end{equation}
where

\begin{equation}
\phi(t)=\sqrt t [C_1 J_{1/3}(\frac{2\sqrt2}{3}t)+C_2 Y_{1/3}(\frac{2\sqrt 2}{3} t)],
\end{equation}
where $J_\nu(t)$ and $Y_\nu(t)$ are the Bessel functions of the first and second kind, respectively .
From the second equation of the system and \eqref{ch}, we obtain

\begin{equation}
\frac{\dot b}{b}=\frac{\dot \phi}{\phi},
\end{equation}
namely:

\begin{equation}
b(t)=\phi(t)+C_3.
\end{equation}

%The general solution \eqref{misks} assume the form 
%\begin{eqnarray}\label{misks}
%u(x,t)&=&-
%\left(\frac{b(t)x^{1-\alpha}}{\Gamma(2-\alpha)}+\frac{2a(t)x^{2-\alpha}}{\Gamma(3-\alpha)}\right)\nonumber\\&=&-
%\left(\frac{b(t)x^{1-\alpha}}{\Gamma(2-\alpha)}+\frac{2a(t)x^{2-\alpha}}{\Gamma(3-\alpha)}\right),
%\end{eqnarray}

\section{Conclusions}
In this paper we have discussed a conceptual link between the
construction of exact solutions to the Schr\"odinger equation with
time-dependent harmonic potential by means of the invariants of the
related Hamiltonian and the problem to construct exact solutions to
nonlinear non homogeneous diffusive equations. Here we consider for
simplicity the construction of solutions by using linear invariants,
but following our ideas there is a bridge between the numerous
results about Time-Dependent Harmonic Oscillator and the solution of
nonlinear diffusive equation with a fixed non-homogeneous term.

\appendix
\section{}
In this section we recall main definitions and properties of
fractional calculus operators Let $\gamma\in \mathbb{R}^{+}$. The
Riemann-Liouville fractional integral is defined by
\begin{equation}
J^{\gamma}_x f(t) =
\frac{1}{\Gamma(\gamma)}\int_0^{x}(x-x')^{\gamma-1}f(x') dx',
\label{riemann-l}
\end{equation}
where
$$\Gamma(\gamma)= \int_0^{+\infty}x^{\gamma-1}e^{-x}dx,$$
is the Euler Gamma function.\\
Note that, by definition, $J^0_x f(t)= f(t)$. \\
Moreover it satisfies the semigroup property, i.e. $J_x^{\alpha}J_x^{\beta} f(t)= J_x^{\alpha+\beta}f(t)$.\\
There are different definitions of fractional derivative (see e.g.
\cite{pod}). In this paper we used the fractional derivatives in
the sense of Caputo, that is
\begin{equation}
D_x^{\gamma}f(x)=   J_x^{m-\gamma} D_x^m f(x)=
\frac{1}{\Gamma(m-\gamma)}\int_0^{x}(x-x')^{m-\gamma-1}\frac{d^m}{d(x')^m}f
(x') \, \mathrm dx', \;\gamma \ne m.
\end{equation}

It is simple to prove the following properties of fractional
derivatives and integrals (see e.g. \cite{Podl}) that will be used
in the analysis:
\begin{align}
&D_x^{\gamma} J_x^{\gamma} f(x)= f(x), \quad \gamma> 0,\\
&J_x^{\gamma} D_x^{\gamma} f(x)= f(x)-\sum_{k=0}^{m-1}f^{(k)}(0)\frac{x^k}{k!}, \qquad \gamma>0, \: x>0,\\
&J_x^{\gamma} x^{\delta}= \frac{\Gamma(\delta+1)}{\Gamma(\delta+\gamma+1)}x^{\delta+\gamma} \qquad \gamma>0, \: \delta>-1, \: t>0,\\
&D_x^{\gamma} x^{\delta}=
\frac{\Gamma(\delta+1)}{\Gamma(\delta-\gamma+1)}x^{\delta-\gamma}
\qquad \gamma>0, \: \delta>-1, \: t>0.
\end{align}

\end{document}